\documentclass[aps, prl, notitlepage, superscriptaddress,12pt]{revtex4-1}

\usepackage[T1]{fontenc}
\usepackage[utf8]{inputenc}
\usepackage{textcomp, amssymb}
\usepackage{graphicx}

\usepackage{amsmath, amssymb, amsfonts}
\usepackage{mathtools}
\usepackage[x11names]{xcolor}
\usepackage[colorlinks=true, citecolor=magenta, linkcolor=blue, urlcolor=blue]{hyperref}
\usepackage{tikz}
\usetikzlibrary{shapes.geometric, arrows, decorations.markings}
\usepackage{enumitem}

\allowdisplaybreaks

\makeatletter

\renewcommand{\@makefnmark}{\hbox{\textsuperscript{\@thefnmark}}}
\makeatother
\usepackage{lineno}

\begin{document}

\title{Hybrid Opto-Electrical Excitation of Spin-Transfer Torque Nano-Oscillators for Efficient Computing}

\author{Felix Oberbauer\footnote[1]{These authors contributed equally to this work.}}
\affiliation{Institut für Physik, Universität Greifswald, 17489 Greifswald, Germany}
\author{Tristan Joachim Winkel\footnotemark[1]}
\affiliation{Institut für Physik, Universität Greifswald, 17489 Greifswald, Germany}
\author{Tim B\"ohnert}
\affiliation{INL - International Iberian Nanotechnology Laboratory, Avenida Mestre José Veiga, s/n, 4715-330 Braga, Portugal}
\author{Marcel S. Claro}
\affiliation{INL - International Iberian Nanotechnology Laboratory, Avenida Mestre José Veiga, s/n, 4715-330 Braga, Portugal}
\author{Luana Benetti}
\affiliation{INL - International Iberian Nanotechnology Laboratory, Avenida Mestre José Veiga, s/n, 4715-330 Braga, Portugal}
\author{Ihsan \c{C}aha}
\affiliation{INL - International Iberian Nanotechnology Laboratory, Avenida Mestre José Veiga, s/n, 4715-330 Braga, Portugal}
\author{Leonard Francis}
\affiliation{INL - International Iberian Nanotechnology Laboratory, Avenida Mestre José Veiga, s/n, 4715-330 Braga, Portugal}
\author{Farshad Moradi}
\affiliation{Electrical and Computer Engineering Department, Aarhus University, 8200 Aarhus, Denmark}
\author{Ricardo Ferreira}
\affiliation{INL - International Iberian Nanotechnology Laboratory, Avenida Mestre José Veiga, s/n, 4715-330 Braga, Portugal}
\author{Markus M\"unzenberg}
\affiliation{Institut für Physik, Universität Greifswald, 17489 Greifswald, Germany}
\author{Tahereh Sadat Parvini\footnote[2]{\href{mailto:Tahereh.Parvini@wmi.badw.de}{Tahereh.Parvini@wmi.badw.de}}}
\affiliation{Institut für Physik, Universität Greifswald, 17489 Greifswald, Germany}
\affiliation{Walther-Meißner-Institut, Bayerische Akademie der Wissenschaften, 85748 Garching, Germany}
\affiliation{Munich Center for Quantum Science and Technology (MCQST), Schellingstr. 4, D-80799 Munich, Germany}

\date{\today}

\begin{abstract}
Neuromorphic computing, inspired by the brain’s parallel and energy-efficient information processing, presents a promising approach for overcoming the limitations of conventional computing architectures. In this study, we investigate spin-transfer torque nano-oscillators (STNOs) as tunable building blocks for neuromorphic applications, leveraging a hybrid excitation scheme that combines AC laser illumination and DC bias currents. Laser-induced thermal gradients generate pulsed thermoelectric voltages ($\mathrm{V_{\text{AC}}}$) via the Tunnel Magneto-Seebeck (TMS) effect, while the addition of a bias current results in a bias-enhanced TMS (bTMS) effect, introducing both $\mathrm{V_{\text{AC}}}$ and a DC voltage component ($\mathrm{V_{\text{DC}}}$). Magnetic field sweeps reveal distinct switching between parallel (P) and antiparallel (AP) magnetization states in both $\mathrm{V_{\text{AC}}}$ and $\mathrm{V_{\text{DC}}}$, suggesting potential for multistate memory applications. In open-circuit conditions, the devices exhibit millivolt-range thermovoltage signals, demonstrating direct compatibility with CMOS technology and providing a promising pathway to simplify scalable and energy-efficient neuromorphic computing architectures by minimizing wiring complexity. Under biased conditions, the devices generate significantly enhanced thermovoltage outputs and exhibit intriguing phenomena, including thermovoltage spikes and double-switching behavior. The observed spikes are correlated with Barkhausen jumps in the DC output voltage, offering fundamental insights into domain dynamics and vortex-core transitions. Beyond their fundamental importance, these spikes bear a striking resemblance to neural spiking behavior, positioning the system as a promising candidate for applications in spiking neural networks and reservoir computing, where complex nonlinear dynamics and memory effects are critical for efficient information processing. These results position the hybrid system of STNOs and vertical-cavity surface-emitting lasers (\mbox{VCSELs}) as a versatile platform for energy-efficient computing, multistate logic, analog computing, advanced signal processing, high-resolution sensing, and neuromorphic computing applications. By bridging fundamental spintronic phenomena with practical applications, this work lays the foundation for next-generation AI technologies and adaptive computational frameworks.

\end{abstract}
\maketitle

\setcounter{footnote}{0}

\section{Introduction}

The rapid advancement of artificial intelligence (AI) has created an urgent need for novel computing architectures that can efficiently handle complex, large-scale tasks. Traditional silicon-based computing, grounded in the von Neumann paradigm, faces fundamental bottlenecks such as limited memory bandwidth, high energy consumption, and the physical separation of processing and storage units \cite{keyes1977physical, keyes2001fundamental, bryant2001limitations}. As AI models grow increasingly sophisticated and data-intensive applications demand greater computational power, these limitations become more pronounced, necessitating the exploration of alternative paradigms such as quantum \cite{dunjko2018machine, mallow2022quantum, ayoade2022artificial, parvini2020, parvini2025} and neuromorphic computing \cite{davies2019benchmarks, markovic2020physics, upadhyay2019emerging, kumar2022dynamical, mohamed2020neuromorphic, 9782767, aimone2022review, markovic2020quantum}.

Neuromorphic computing aims to emulate the highly parallel, adaptive, and energy-efficient processing of biological neural networks. Unlike conventional digital processors, neuromorphic systems operate through dynamic state transitions in artificial neurons and synapses, enabling cognitive computing, real-time edge processing, and intelligent sensing applications \cite{sung2022simultaneous, du2021synaptic, chen2023biological, Zins2023, huang2021memristive}. A promising avenue for hardware implementation is spintronic devices \cite{grollier2020neuromorphic, torrejon2017neuromorphic, zhou2021prospect}, particularly spin-transfer torque nano-oscillators (STNOs) \cite{sadat2023enhancing, jha2023interface, 8268505, liu2022compensated, jiang2024spin, yu2020nondestructive}. These nanoscale devices leverage spin-polarized currents to generate high-frequency oscillations, exhibiting rich nonlinear dynamics that can be exploited for neuromorphic information processing \cite{7280306, zhang2021spin, kurenkov2019artificial, vincent2015spin}. Among various STNO designs, vortex-based STNOs offer enhanced stability and robustness due to their localized magnetic vortex structures, making them a promising candidate for neuromorphic applications, although challenges in large-scale integration remain \cite{jenkins2021electrical, imai2022input, suess2018topologically, kammerer2011magnetic}. While recent advances have demonstrated neuromorphic computing capabilities in STNOs, primarily through frequency and phase encoding, significant challenges persist. Efficient thermal management, scalable architectures, and advanced signal processing techniques are crucial for unlocking their full computational potential. In this study, we explore the dynamic response of dimensionally optimized STNOs under laser-induced thermal gradients \cite{chen2017all, boehnke2017large, xu2016origins, yang2018magnetic, lin2012giant}. We introduce a hybrid excitation scheme combining optical and electrical inputs to enhance tunability and control. A modulated diode laser (1 kHz on-off frequency) periodically heats the tunnel barrier, generating a pulsed thermoelectric voltage ($V_{\text{AC}}$) via the Tunnel Magneto-Seebeck (TMS) effect, which arises from temperature-dependent asymmetries in electronic states. Concurrently, an applied bias current induces a DC voltage component ($V_{\text{DC}}$), forming the baseline of the output pulse. By systematically varying external magnetic fields, we disentangle the contributions of these components, providing insights into the interplay between thermal and electrical control in STNO-based neuromorphic architectures.

\begin{figure*}[t!]
  \centering
  \includegraphics[width=0.99\textwidth]{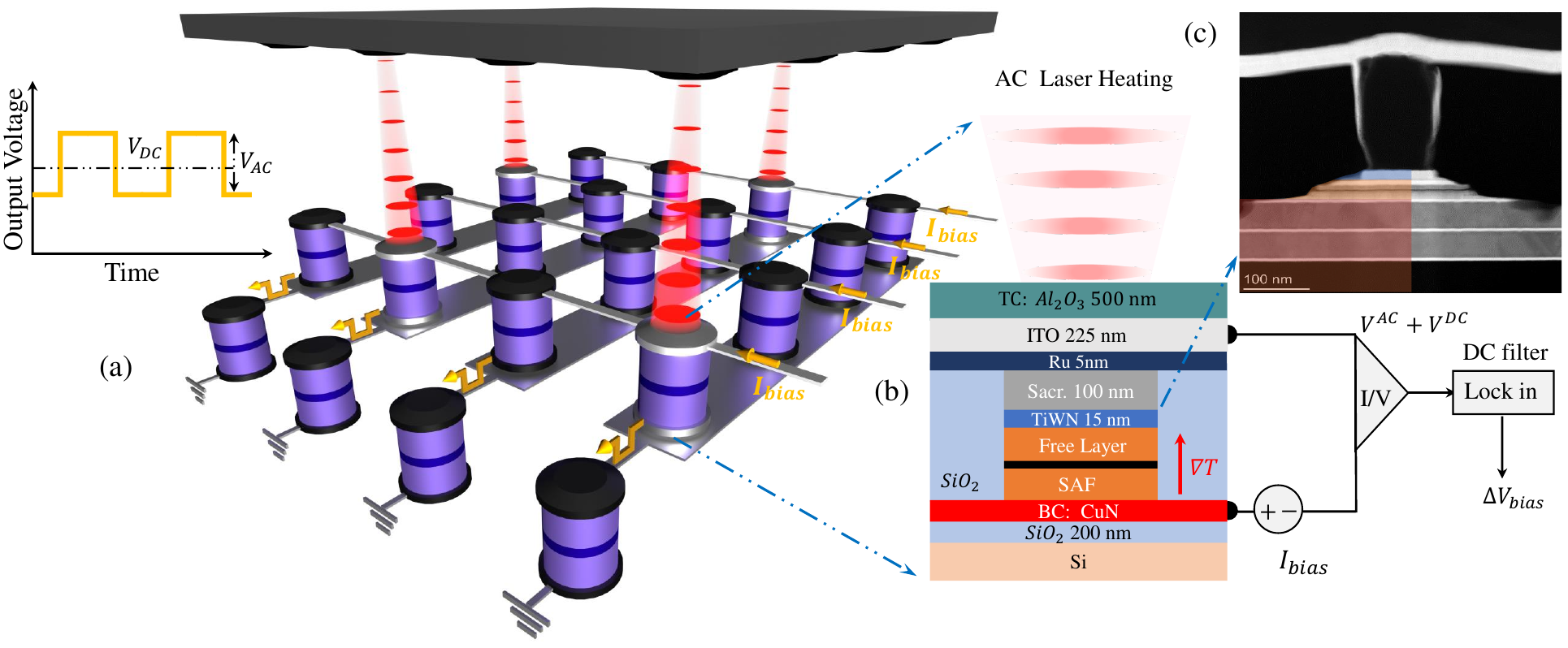}
  \caption{(a) Schematic representation of a hybrid photonic-spintronic neuromorphic system. A VCSEL array provides spatially selective optical excitation to an STNO array, enabling independent control and dynamic modulation of individual devices. (b) Schematic of the STNO configuration and the fundamental principles of the Tunnel Magneto-Seebeck (TMS) and bias-dependent TMS (bTMS) experiments. The electrical setup includes a lock-in amplifier, which isolates the AC signal generated by laser-induced heating. (c) High-angle annular dark-field scanning transmission electron microscopy (HAADF-STEM) image of an isolated magnetic tunnel junction (MTJ) nanopillar.}
\label{Fig1}
\end{figure*}

The overarching vision of this work is captured in Fig.~\ref{Fig1}(a), which illustrates a neuromorphic chip integrating spintronic devices with the optical precision of vertical-cavity surface-emitting lasers (VCSELs) for localized optical excitation. This architecture utilizes programmable VCSEL arrays to selectively activate STNOs, each exhibiting unique thermoelectric responses to laser-induced heating. Such device-specific responses enable spatial and temporal information encoding, supporting parallel processing and reconfigurable architectures. By dynamically modulating laser activation patterns, the system achieves hierarchical computations, addressing tasks such as spiking neural networks \cite{ghosh2009spiking,hu2024advancing,yang2018real}, reservoir computing, and addressable multistate memory \cite{rzeszut2022multi,shen2024multilevel} arrays. Furthermore, the pulsed thermoelectric outputs from the first layer can serve as inputs for a second layer, enabling sequential and hierarchical information processing. In this configuration, the second layer operates without additional biasing, relying solely on the output signals of the first layer. This layered design enhances energy efficiency while supporting adaptive computations, where each layer refines or transforms the information passed from the previous one. This hierarchical approach mirrors biological neural networks and opens pathways to modular, scalable architectures for advanced AI applications, including spatiotemporal processing, real-time decision-making, multistage logic operations, and high-precision analog computing.

\section{Device Features and Laser Irradiation Setup}

An array of spin-transfer torque nano-oscillators (STNOs) was fabricated, featuring magnetic tunnel junction (MTJ) nanopillars with cross-sectional geometries ranging from circular to elliptical shapes of varying radii (Fig.~\ref{Fig1}(b)). Each nanopillar was embedded within a SiO$_2$ dielectric layer and engineered with a multilayered structure optimized for performance, ensuring uniformity across all devices. The free layer comprised CoFe$_{40}$B$_{20}$ (2)/Ta (0.21)/Ni$_{80}$Fe$_{20}$ (7)/Ta (10)/Ru (7), while a sacrificial layer of TiWN (15) and AlSiCu (100) facilitated the connection between the magnetic layers and the top contact. Magnetic stability was achieved by integrating a synthetic antiferromagnet (SAF) layer consisting of Ta (5)/Ru (5)/IrMn (6)/CoFe$_{30}$ (2)/Ru (0.7)/CoFe$_{40}$B$_{20}$ (2.6). The bottom contact was constructed as Ta (5)/CuN (50)/Ta (5)/CuN (50)/Ta (5)/Ru (5). All thicknesses are given in nanometers. To enhance optical accessibility, an indium tin oxide (ITO) top contact was employed \cite{jin2016optically, maniyara2021highly, farhan2013electrical}. 

To investigate the influence of laser-induced thermal gradients on STNO dynamics, we employed a modulated diode laser (638 nm wavelength, 1 kHz modulation frequency) focused to a $7~\mu$m diameter spot. Calibrated laser powers of 30, 60, 90, 120, and 150 mW at the source corresponded to effective powers of 15, 28, 41, 54, and 67 mW incident on the sample surface, accounting for optical losses. Periodic laser illumination generated a temperature gradient across the MTJ tunnel barrier, inducing a pulsed thermoelectric voltage ($\mathrm{V_\text{AC}}$). Additionally, a bias current was applied to modulate transport properties, introducing a DC voltage component ($\mathrm{V_\text{DC}}$). Under open-circuit conditions ($\mathrm{I_\text{Bias}=0}$), $\mathrm{V_\text{DC}=0}$. The AC and DC voltage components were separated and independently measured using dedicated electronic circuitry (Fig.~\ref{Fig1}). Magnetic field sweeps revealed hysteretic switching behavior in both \(\mathrm{V_{\text{AC}}}\) and \(\mathrm{V_{\text{DC}}}\), with distinct responses for parallel (P) and antiparallel (AP) magnetization states. This tunable multistate behavior suggests promising applications in memory and logic devices, where the interplay between thermal and electrical excitation could enable novel information processing functionalities.

\section {Experimental results and discussion}
\subsection{Thermoelectric Response of STNOs with Circular Nanopillars}

The \(\mathrm{V_{\text{AC}}}\), representing the Seebeck voltage, arises from thermally excited charge carriers due to the temperature gradient across the magnetic tunnel junction (MTJ) nanopillars \cite{uchida2008observation, czerner2011spin, walter2011seebeck, gravier2006thermodynamic}. This thermoelectric response is inherently linked to the asymmetry of electronic states around the Fermi level in the ferromagnetic layers \cite{walter2011seebeck, schmidt2018boltzmann, boehnke2017large}. Specifically, the difference in the spin-up and spin-down electron densities at the Fermi level leads to a spin-dependent Seebeck effect, making the Seebeck voltage highly sensitive to the nanopillar’s magnetic configuration. As the external magnetic field is swept, \(\mathrm{V_{\text{AC}}}\) exhibits hysteretic switching between parallel (P) and antiparallel (AP) magnetization states, a hallmark of the tunnel magneto-Seebeck (TMS) effect. This effect provides a robust readout mechanism for spintronic logic operations.

\begin{figure*}[t!]
  \centering
  \includegraphics[width=0.95\textwidth]{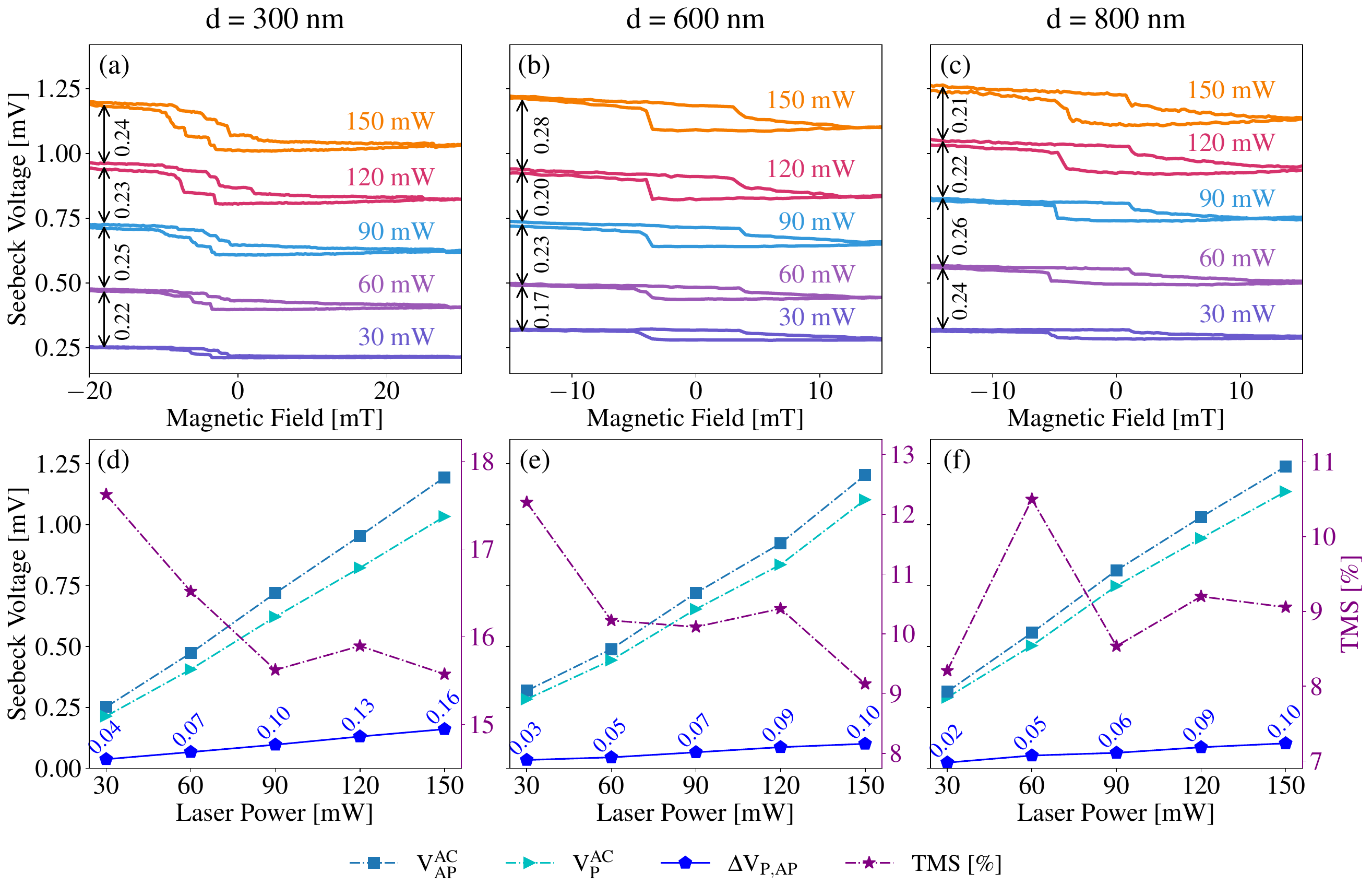}
  \caption{Seebeck voltage as a function of magnetic field for varying laser powers in STNOs with nanopillar diameters of 300 nm ((a) and (d)), 600 nm ((b) and (e)), and 800 nm ((c) and (f)).}
\label{Fig2}
\end{figure*} 
To systematically investigate the TMS effect in circular STNOs, we conducted magnetic field sweeps under open-circuit conditions across different laser power levels. The resulting hysteresis loops reveal distinct AC Seebeck voltages for P and AP states (V\(\mathrm{V_{P}^{AC}}\) and \(\mathrm{V_{AP}^{AC}}\)). Notably, the magnetic switching fields observed in \(\mathrm{V_{AC}}\) closely match those in magnetoresistance measurements of biased STNOs, confirming the correlation between thermoelectric and spintronic transport properties. The thermoelectric voltage (\(\mathrm{V_{P}^{AC}}\) and \(\mathrm{V_{AP}^{AC}}\)) is related to the Seebeck coefficient~(\(\mathrm{S}\)) and the applied temperature gradient (\(\mathrm{\Delta T}\)) as $\mathrm{V_{P,AP}^{AC} = S_{P,AP} \Delta T}$ where \(\mathrm{S_P}\) and \(\mathrm{S_{AP}}\) are the Seebeck coefficients for the parallel and antiparallel magnetization states, respectively. Analogous to the tunnel magnetoresistance (TMR) ratio, we define the TMS effect ratio as:
\begin{equation}
    \mathrm{TMS = \frac{S_{AP} - S_{P}}{\min(|S_{AP}|, |S_{P}|)} \equiv \frac{V_{AP} - V_{P}}{\min(|V_{AP}|, |V_{P}|)}}.
\end{equation}

As the $\mathrm{S_{P,AP}}$, along with their corresponding voltages, can take negative values, division by the minimum absolute value ensures a meaningful ratio. Fig.\ref{Fig2}(a–c) present the measured Seebeck voltage as a function of applied magnetic field for STNOs with 300 nm, 600 nm, and 800 nm nanopillars under varying laser powers. The extracted values of (\(\mathrm{V_{P}^{AC}}\)) and (\(\mathrm{V_{AP}^{AC}}\)), their difference (\(\Delta \mathrm{V_{P,AP}} = \mathrm{V_{AP}^{AC}} - \mathrm{V_{P}^{AC}}\)) and the resulting TMS ratio are shown in Fig.\ref{Fig2}(d–f) as a function of laser power. Devices with 300 nm nanopillars exhibit a larger $\Delta \mathrm{V_{P,AP}}$ and higher TMS ratios, likely due to steeper thermal gradients, reduced carrier scattering, and enhanced asymmetry in the electronic density of states at the nanoscale. The millivolt-range Seebeck voltages are CMOS-compatible, eliminating the need for biasing wires and facilitating scalable integration into neuromorphic systems. The observed linear dependence of \(\mathrm{V_P^{AC}}\), \(\mathrm{V_{AP}^{AC}}\) and $\Delta \mathrm{V_{P,AP}}$ on laser power, consistent with $\mathrm{V_{P,AP}^{AC} = S_{P,AP} \Delta T}$, underscores the precise tunability of the thermoelectric response. This tunability offers significant potential for neuromorphic computing, where it can emulate synaptic weights, and for analog computing, where its continuous, linear behavior is ideal for approximating neuron activation functions. Additionally, its sensitivity to temperature and magnetic fields highlights applications in high-resolution sensing.

To extend the functionality of the TMS effect and investigate its potential for advanced applications, we introduce the bias-enhanced Tunnel Magneto-Seebeck (bTMS) effect \cite{boehnke2015off, kuschel2019tunnel}, which provides enhanced precision in controlling transport properties. Through the analysis of voltage variations using the Onsager transport equation \cite{onsager1931reciprocalI, onsager1931reciprocalII}, the voltage difference between laser on and off states under conditions with and without laser-induced heating in the STNO can be expressed as:
\begin{equation}
    \mathrm{\Delta V_{P,AP}=V_{on}-V_{off}=S_{P,AP}\Delta T+\left(R_{P,AP}-\Delta R_{P,AP}\right)I-R_{P,AP}I},
\end{equation}
where $\mathrm{R}$ is the resistance of the non-heated MTJ, and $\mathrm{R-\Delta R}$ is the resistance when the STNO is heated. To quantify the on/off switching of the measured AC voltage induced by magnetization reversal, we defined a bTMS ratio as
\begin{equation}
 \mathrm{bTMS=\frac{\Delta V_{AP}-\Delta V_{P}}{min\left(\left|\Delta V_{P}\right|,\left|\Delta V_{AP}\right|\right)}}.
\end{equation}

\begin{figure*}[t!]
  \centering
  \includegraphics[width=0.95\textwidth]{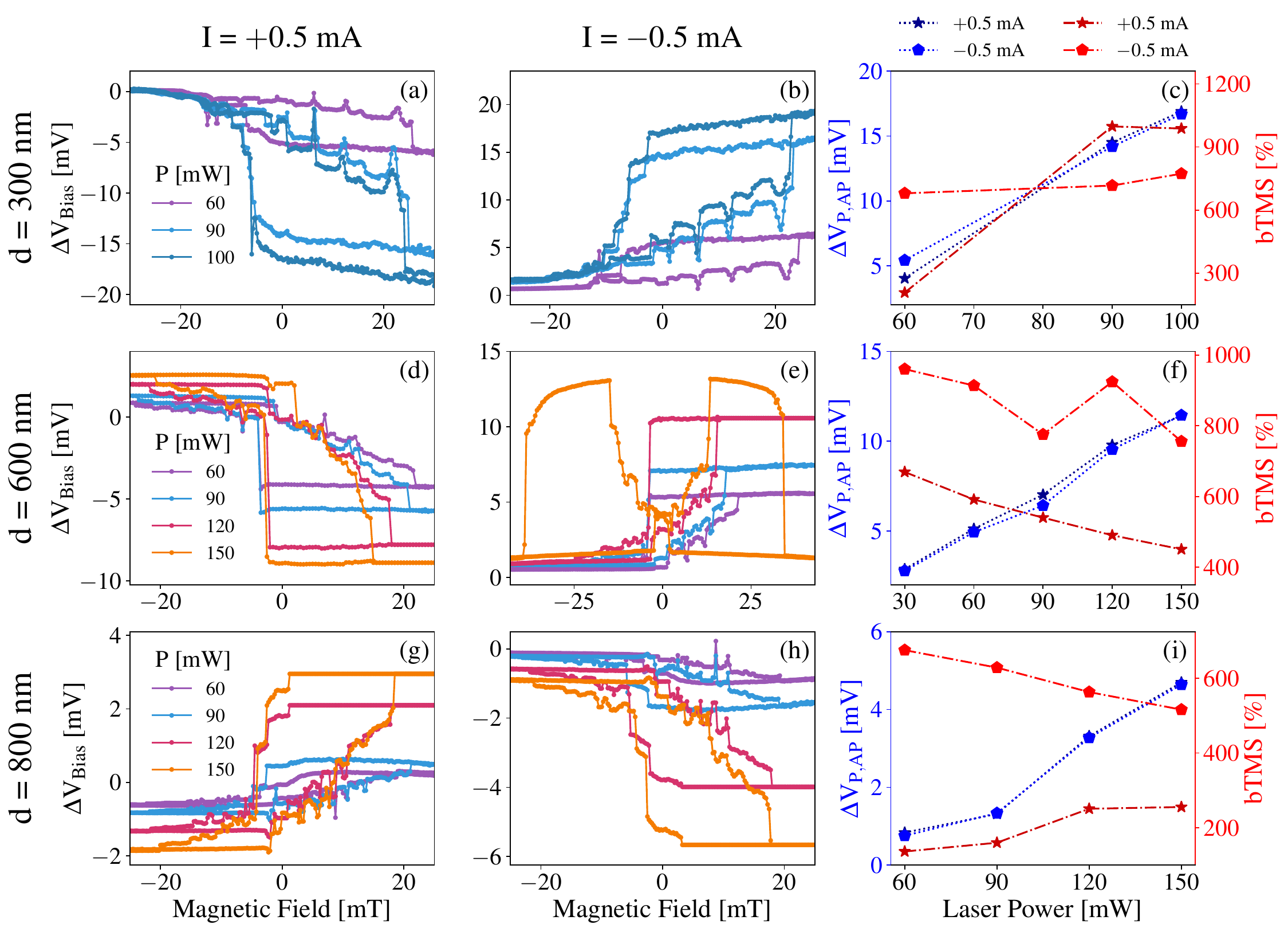}
  \caption{The voltage $\mathrm{\Delta V_{Bias}}$ as a function of the magnetic field for various laser powers at a bias current of +0.5 mA ((a), (d), (g)), at a bias current of -0.5 mA ((b), (e), (h)) and the resulting bTMS ratios and voltage differences between P and AP states ((c), (f), and (i)).}
\label{fig3}
\end{figure*} 

The AC voltage as a function of the magnetic field for devices with nanopillar diameters of 300 nm ((a), (b)), 600 nm ((d), (e)), and 800 nm ((g), (h)) is presented in Fig. \ref{fig3}. Measurements were conducted under positive bias currents of 0.5 mA ((a), (d), (g)) and negative bias currents of -0.5 mA ((b), (e), (h)) for varying laser powers. These results revealed a giant thermoelectric voltage with pronounced contrast between P and AP states, significantly enhancing the signal-to-noise ratio and readout precision.

As shown in SM \cite{SM}, the tunneling magnetoresistance (TMR) curves of these devices exhibit discrete step-like features, which, depending on the device dimensions, can be attributed to different phenomena. For smaller devices (d = 300 nm), these steps originate from Barkhausen jumps due to domain wall motion overcoming pinning sites \cite{kuepferling2015vortex, bohn2018playing}.In larger devices (600–800 nm), the free layer stabilizes a vortex magnetization state. This flux-closure configuration minimizes magnetostatic energy and becomes energetically favorable once the device dimensions exceed a critical threshold \cite{yoo2012radial}. In such cases, the resistance steps are linked to jumps between local pinning sites of the vortex core, residing at the center of the vortex magnetization state, as detailed in \cite{jenkins2024impact}. Remarkably, the thermovoltage output exhibits sharp spikes at the exact magnetic field values where resistance steps occur, demonstrating extreme sensitivity to abrupt magnetization changes. This nonlinear response provides insights into vortex and domain wall dynamics and suggests applications in ultra-sensitive magnetic field sensors. Additionally, the discrete thermovoltage jumps could enable multi-state synaptic devices, where intermediate magnetization/vortex states encode information with high fidelity. Furthermore, the observed nonlinear magnetization dynamics can serve as a robust physical reservoir for reservoir computing \cite{riou2021reservoir}. By mapping complex input signals, such as magnetic field variations or thermal gradients, into a high-dimensional space, these dynamics enable advanced computational tasks like chaotic time-series prediction and real-time classification \cite{shreya2023granular}. While this study does not address time-resolved behavior, the sharp thermovoltage spikes indicate the potential for high-resolution sensing and unconventional computing paradigms, warranting further exploration.

\begin{figure}[t!]
  \hspace*{1.5cm}
  \includegraphics[width=0.78\textwidth]{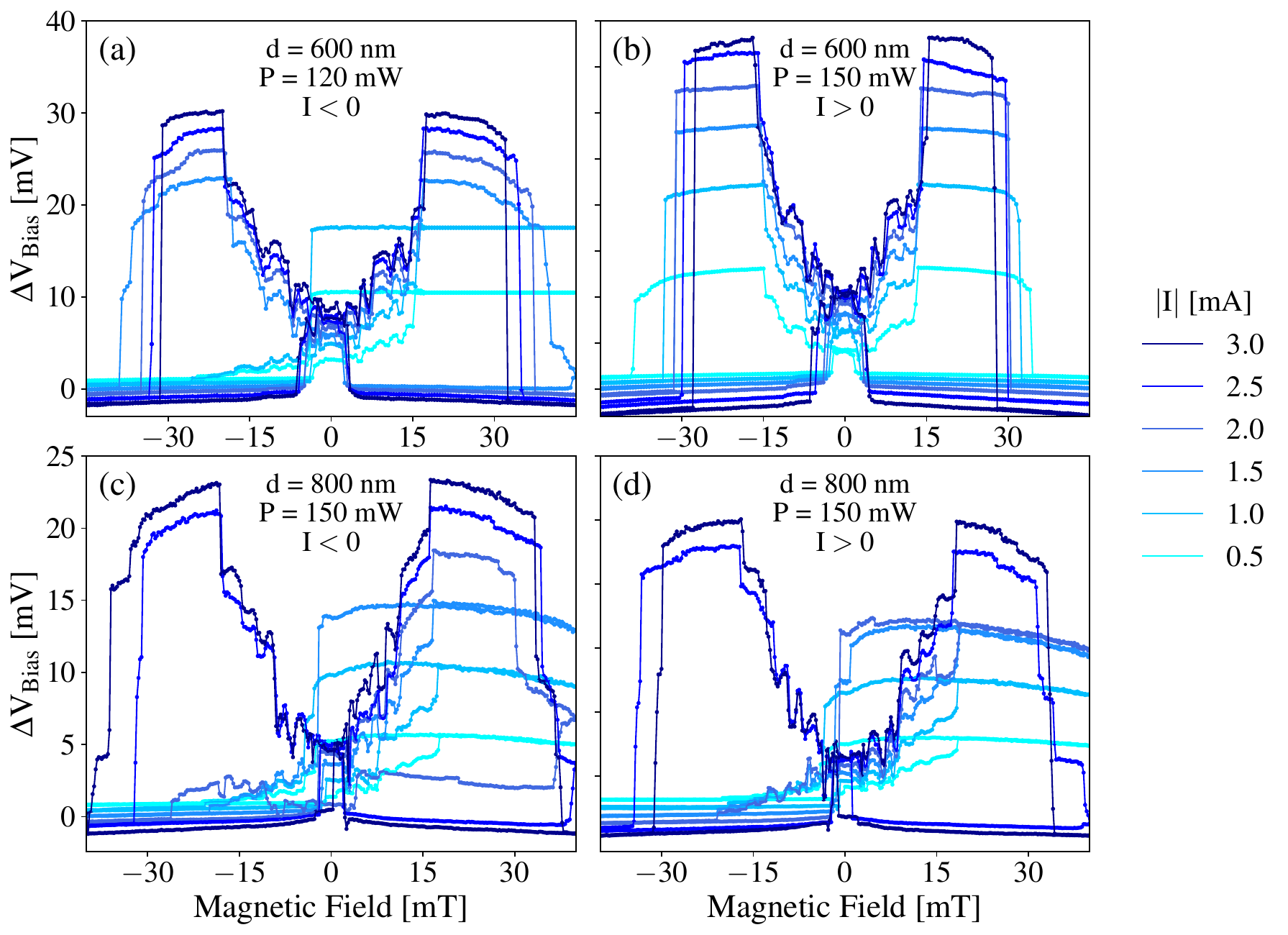}
  \caption{The measured thermovoltage, $\mathrm{\Delta V_{\text{Bias}}}$, as a function of the external magnetic field for a nanopillar with $\mathrm{d} = 600~\mathrm{nm}$ ((a) and (b)) and $\mathrm{d} = 800~\mathrm{nm}$ ((c) and (d)), at varying laser power and bias current.}
\label{fig4}
\end{figure}

As shown in Fig.~\ref{fig3}(e), the 600~nm STNO exhibits an unconventional double-switching behavior at a threshold bias current of \(\mathrm{I} = -0.5~\mathrm{mA}\) and laser power of \(\mathrm{P} = 150~\mathrm{mW}\). This phenomenon is consistently observed across all devices when both the bias current and laser power exceed specific threshold values. Results for the 600~nm and 800~nm devices are presented in Fig.~\ref{fig4}, while the corresponding behavior for the 300~nm device is detailed in the SM \cite{SM}. The precise origin of this double-switching behavior remains an open question. One possible explanation is that at elevated laser powers and bias currents, the combined effects of thermal gradients and spin-transfer torque (STT) drive the system into a nonlinear regime \cite{zeng2013spin, romera2015non, uzunova2023nonlinear}, leading to complex switching characterized by multiple magnetization transitions. Another possibility is that under extreme experimental conditions, the fixed magnetic layer becomes unstable, resulting in precessional motion or partial magnetization reversal, which in turn alters the overall device response. These scenarios emphasize the need for further investigation, particularly time-resolved magnetization measurements, to fully understand the underlying mechanisms and their implications for spintronic applications. Such behavior could also have implications for neuromorphic computing, where it may emulate synaptic responses for adaptive learning systems.

\subsection{Thermoelectric Effects in STNOs with Elliptical Nanopillars}

\begin{figure}[t!]
  \centering
  \includegraphics[width=0.7\textwidth]{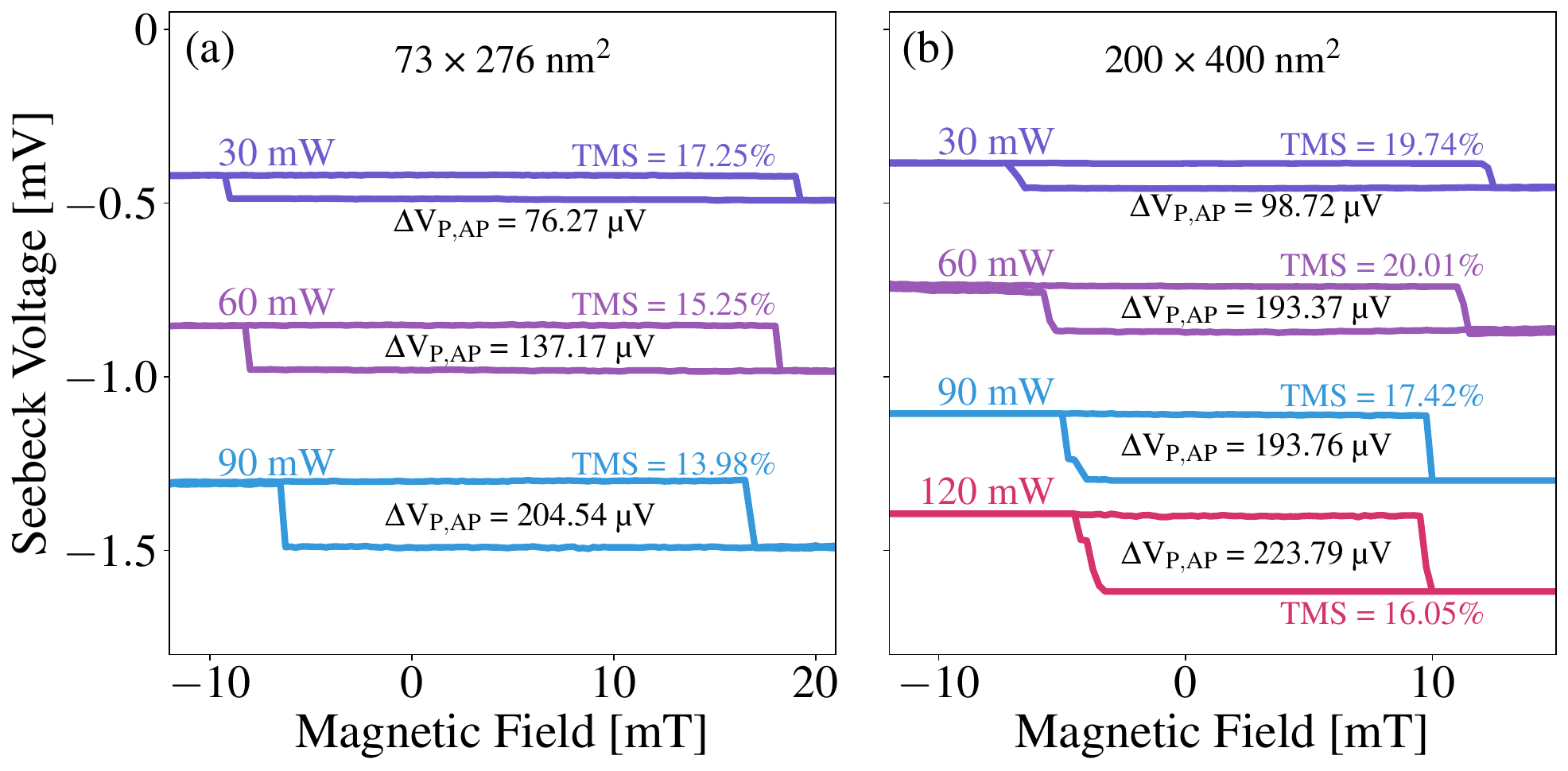}
  \caption{Magnetic field and laser power dependence of the tunnel thermovoltage in STNOs with elliptical nanopillar: (a) $\mathrm {73\times276 \, \, nm^2}$ and (b) $\mathrm {200\times400  \, \, nm^2}$. Related TMS ratio values are presented adjacent to each curve.}
\label{fig5}
\end{figure}
STNOs with elliptical cross-sections exhibit in-plane magnetic anisotropy, stabilizing two energetically favorable antiparallel magnetization states along the major axis. This intrinsic anisotropy enables binary encoding (0 and 1), making these structures highly relevant for spintronic memory and logic applications. TMS measurements for two STNOs with dimensions of $\mathrm {73\times276 \, \, nm^2}$ (eccentricity e = 0.96) and $\mathrm {200\times400  \, \, nm^2}$ (e = 0.87) are presented in Fig.~\ref{fig5}. The smaller elliptical nanopillar ($\mathrm{73\times276 \, \, nm^2}$) exhibits greater shape anisotropy and a stronger demagnetizing field, resulting in a higher energy barrier for magnetization switching. Consequently, a stronger external magnetic field is required to overcome this barrier, as evidenced by the increased switching field compared to the larger one (see Fig. \ref{fig5}). Increasing laser power reduces the switching field by lowering the energy barrier and decreasing the saturation magnetization, thereby weakening the demagnetizing field. Additionally, these elliptical STNOs, despite their smaller size relative to the circular 300 nm devices, exhibit a larger voltage difference ($\mathrm{\Delta V_{P,AP}}$),  indicating a higher contrast between the parallel (P) and antiparallel (AP) states.

\begin{figure}[h!]
  \centering
  \includegraphics[width=0.90\textwidth]{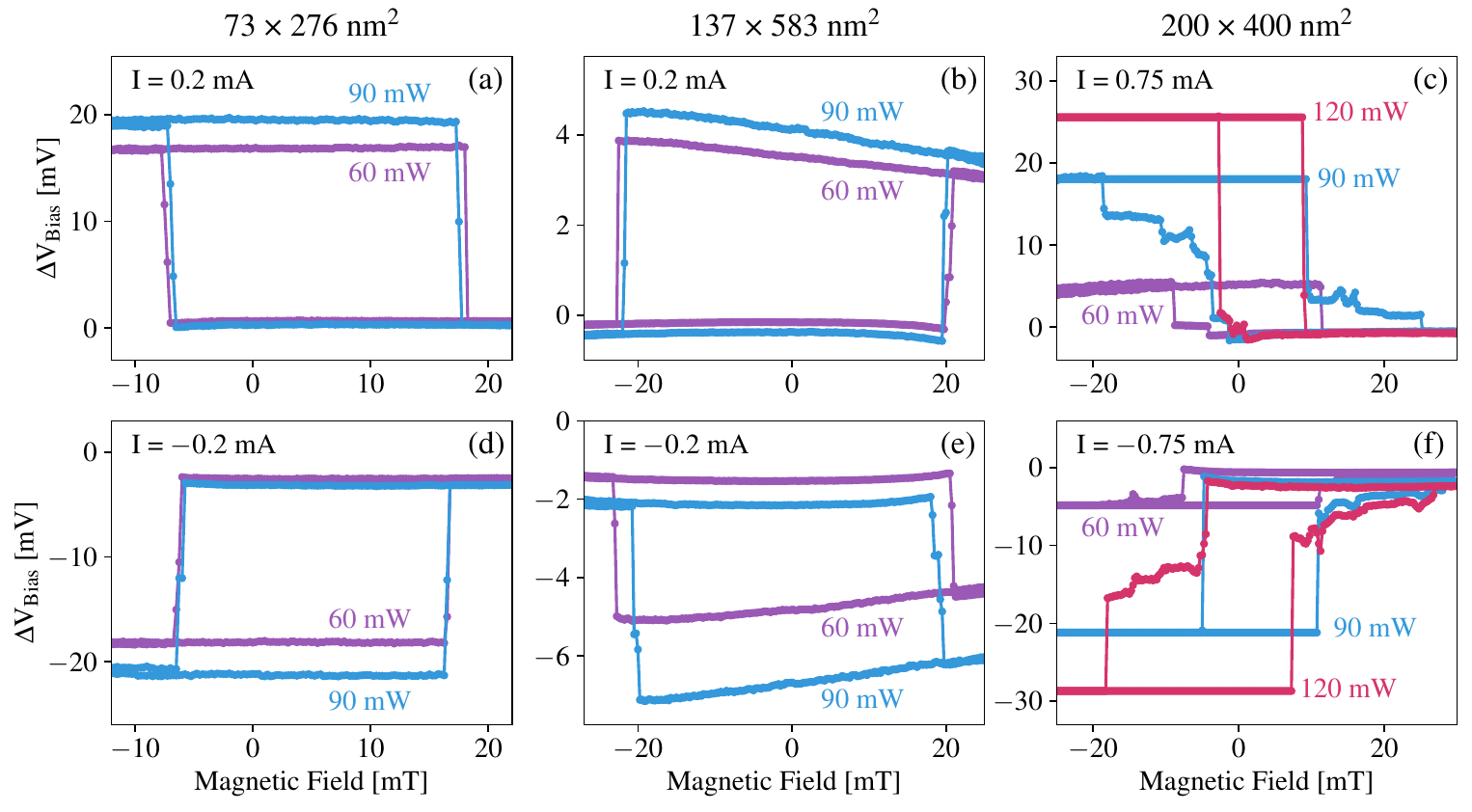}
  \caption{$\mathrm{\Delta V_{Bias}}$ as a function of the magnetic field for varying laser powers and bias currents in STNOs with elliptical nanopillars: (a) and (d) $\mathrm {73\times276  \, \, nm^2}$ (e = 0.96), (b) and (e) $\mathrm {137\times583  \, \, nm^2}$ (e = 0.97), and (c) and (f) $\mathrm {200\times400  \, \, nm^2}$ (e = 0.87).}
\label{fig6}
\end{figure}

$\mathrm{\Delta V_{Bias}}$ as a function of the magnetic field for varying laser powers and bias currents is shown in Fig. \ref{fig6} for elliptical STNOs with different eccentricities. The $\mathrm {200\times400  \, \, nm^2}$ STNO, with a smaller eccentricity, demonstrates lower magnetic field switching thresholds and higher thermovoltage output compared to devices with greater eccentricities. Interestingly, the measured $\mathrm{\Delta V_{Bias}}$ displays small jumps, reminiscent of features commonly associated with vortex states. This suggests that elliptical nanopillars with smaller eccentricities may support vortex formation under specific laser power, potentially influencing device performance and stability. Our results demonstrate that arrays of STNOs with identical layer configurations but diverse cross-sectional geometries and dimensions on a single chip can generate a wide range of responses, which is critical for advanced computing applications.

\section{\textbf{Conclusion}}

In summary, we have fabricated and optimized an array of STNOs with both circular and elliptical cross-sections. A key aspect of this work is the implementation of a hybrid excitation scheme, enabling independent or simultaneous activation via AC laser illumination and DC bias current. Laser illumination alone induced substantial millivolt-range thermoelectric voltages via the TMS effect, demonstrating CMOS compatibility. This breakthrough paves the way for simplified, energy-efficient neuromorphic chip architectures with reduced wiring complexity. Incorporating a bias current further enhances the system’s response, leading to a bias-enhanced TMS (bTMS) effect with significantly increased thermoelectric outputs. The observed thermovoltage spikes provide direct insight into vortex state transitions, revealing exceptional sensitivity to magnetization changes, vortex core motion, and pinning site interactions—critical phenomena in STNO dynamics. Additionally, the double-switching behavior observed under specific biasing and laser power conditions suggests the presence of nonlinear magnetization dynamics or instabilities in the fixed layer, requiring further investigation to fully elucidate its origin.

Beyond their fundamental significance, the pulsed nature of the output voltage in both TMS and bTMS regimes resembles the spiking behavior of biological neurons, suggesting potential applications in systems inspired by neuromorphic computing. This behavior enables the integration of large arrays of nanopillars, each optically excited at distinct frequencies and sharing a common electrical output, creating an efficient frequency-multiplexing scheme. By employing frequency-selective devices in subsequent processing layers, this approach could reduce hardware complexity and enhance scalability \cite{leroux2021hardware}. Furthermore, the combination of optical excitation and frequency-encoded spintronic signal processing offers opportunities for hybrid photonic-spintronic architectures, which may contribute to advancements in adaptive computing and signal processing. Future time-domain studies will be crucial to unravel the transient dynamics of pulsed thermovoltage and double-switching behavior, providing further insights into their underlying mechanisms and potential applications in neuromorphic systems, high-resolution sensing, and multistate memory storage. By leveraging their unique thermal, magnetic, and nonlinear dynamical properties, hybrid laser-STNO systems hold significant promise for bridging the gap between fundamental spintronic physics and next-generation computing technologies.

\section{\textbf{Acknowledgment}}

This project was supported by funding from the European Union’s Horizon 2020 research and innovation program under grant agreement No. 899559 (SpinAge). We thank J. Walowski, T. Ahlgrimm, L. Vollroth, and M. Kohlmann for their support with the experimental setup and technical discussions.

\bibliography{1_main.bib}
\clearpage
\end{document}


\title{Supplementary Material: Hybrid Opto-Electrical Excitation of Spin-Transfer Torque Nano-Oscillators for Efficient Computing}

\author{Felix Oberbauer\footnote[1]{These authors contributed equally to this work.}}
\affiliation{Institut für Physik, Universität Greifswald, 17489 Greifswald, Germany}
\author{Tristan Joachim Winkel\footnotemark[1]}
\affiliation{Institut für Physik, Universität Greifswald, 17489 Greifswald, Germany}
\author{Tim B\"ohnert}
\affiliation{INL - International Iberian Nanotechnology Laboratory, Avenida Mestre José Veiga, s/n, 4715-330 Braga, Portugal}
\author{Marcel S. Claro}
\affiliation{INL - International Iberian Nanotechnology Laboratory, Avenida Mestre José Veiga, s/n, 4715-330 Braga, Portugal}
\author{Luana Benetti}
\affiliation{INL - International Iberian Nanotechnology Laboratory, Avenida Mestre José Veiga, s/n, 4715-330 Braga, Portugal}
\author{Ihsan \c{C}aha}
\affiliation{INL - International Iberian Nanotechnology Laboratory, Avenida Mestre José Veiga, s/n, 4715-330 Braga, Portugal}
\author{Leonard Francis}
\affiliation{INL - International Iberian Nanotechnology Laboratory, Avenida Mestre José Veiga, s/n, 4715-330 Braga, Portugal}
\author{Farshad Moradi}
\affiliation{Electrical and Computer Engineering Department, Aarhus University, 8200 Aarhus, Denmark}
\author{Ricardo Ferreira}
\affiliation{INL - International Iberian Nanotechnology Laboratory, Avenida Mestre José Veiga, s/n, 4715-330 Braga, Portugal}
\author{Markus M\"unzenberg}
\affiliation{Institut für Physik, Universität Greifswald, 17489 Greifswald, Germany}

\author{Tahereh Sadat Parvini\footnote[2]{\href{mailto:Tahereh.Parvini@wmi.badw.de}{Tahereh.Parvini@wmi.badw.de}}}
\affiliation{Institut für Physik, Universität Greifswald, 17489 Greifswald, Germany}
\affiliation{Walther-Meißner-Institut, Bayerische Akademie der Wissenschaften, 85748 Garching, Germany}


\maketitle
\setcounter{footnote}{0}

\section{Bias-Enhanced TMS}

\begin{figure*}[t!]
  \centering
  \includegraphics[width=0.75\textwidth]{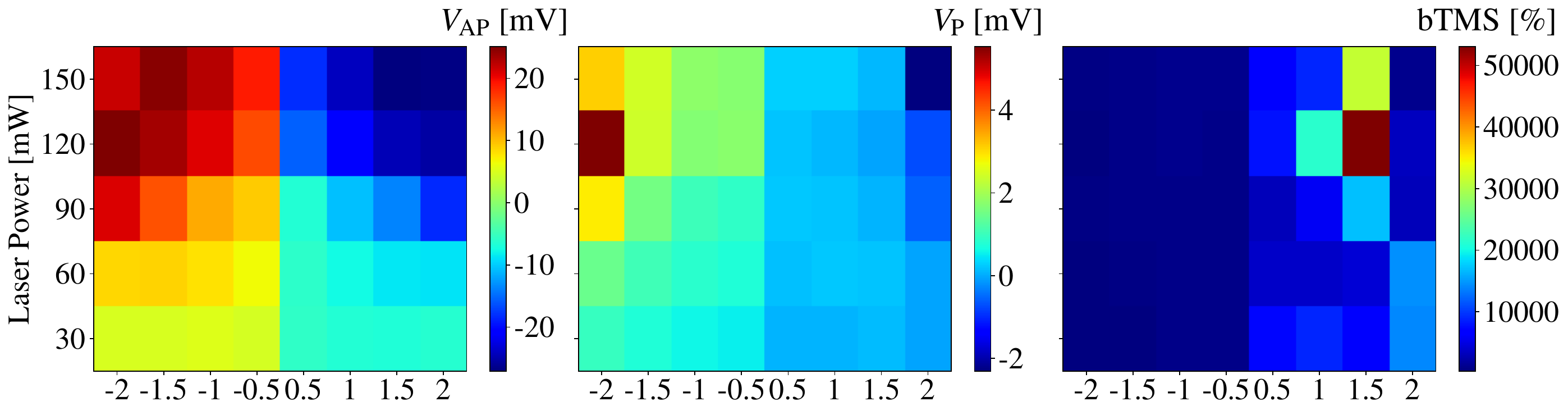}\\
  \includegraphics[width=0.75\textwidth]{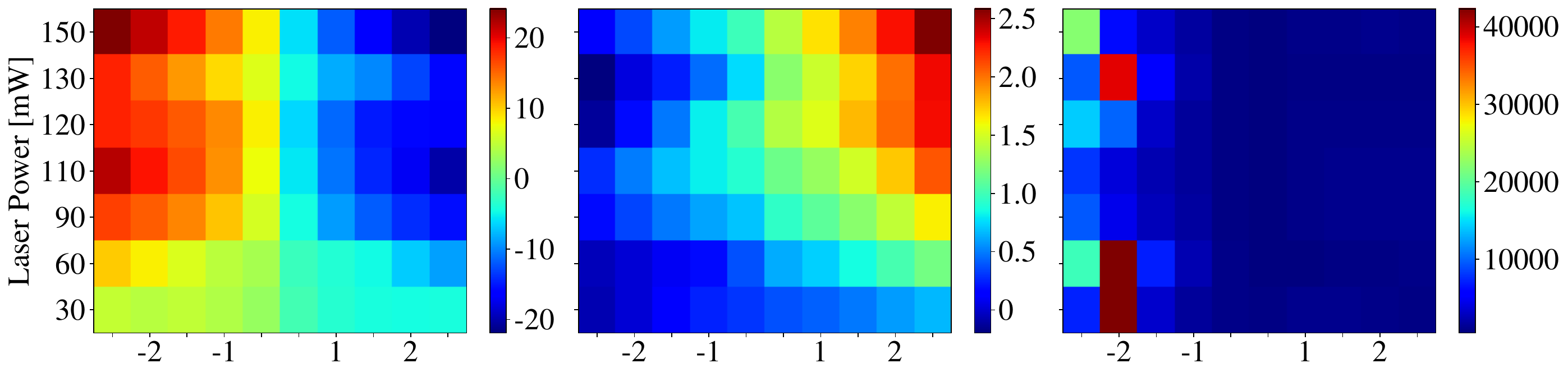}\\
  \includegraphics[width=0.75\textwidth]{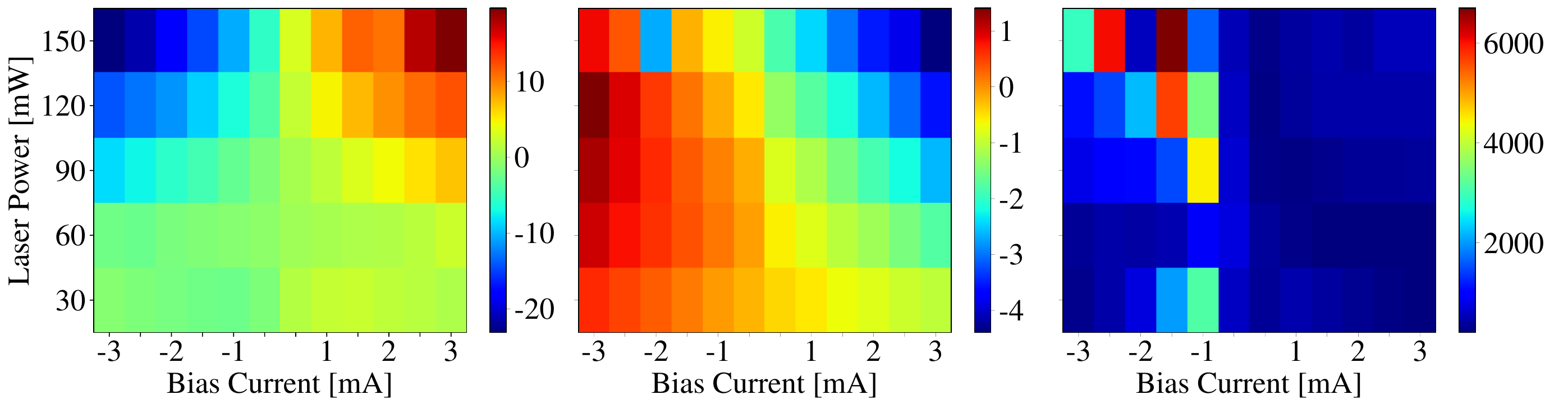}
  \caption{Measured voltage signals at P- and AP-states, along with the bTMS ratio, as functions of laser power and bias current for devices with nanopillar diameters: 300 nm (first row), 600 nm (second row), and 800 nm (third row).}
\label{Fig1SM}
\end{figure*}

We optimized spin-transfer torque nano-oscillators (STNOs) for enhanced performance in neuromorphic computing \cite{sadat2023enhancing, oberbauer2025hybrid}. The laser-induced temperature gradient across the MgO barrier in the magnetic tunnel junction (MTJ) drives the tunnel magneto-Seebeck (TMS) effect, enabling precise charge transport modulation, where the laser acts solely as a heat source without direct light-matter interaction \cite{liu2023lightning, parvini2025, du2023deterministic, hamidi2013, parvini2015}. Importantly, bTMS is introduced to precisely control charge and spin transport across the magnetic tunnel junction by leveraging the interplay between bias current and laser-induced thermal gradients. The measured voltage signals at the P- and AP-states, along with the corresponding bTMS ratios, are presented in Fig. \ref{Fig1SM} as functions of laser power and bias current for devices with nanopillar diameters of 300 nm, 600 nm, and 800 nm. The first row corresponds to the 300 nm nanopillar, the second row to the 600 nm nanopillar, and the third row to the 800 nm nanopillar. Across all devices, we observe a giant thermovoltage in the millivolt range. The STNO with a 300 nm nanopillar cross-section generates a higher output voltage than larger devices (600 nm and 800 nm), primarily due to its distinct magnetic configuration. Although larger nanopillars exhibit vortex states, the 300 nm devices maintain a simpler domain structure that prevents vortex formation. This simplified configuration enhances spin polarization, minimizes magnetic noise, and facilitates higher current densities, collectively resulting in higher output voltages in bTMS measurements.

\section{Double Switching and Thermovoltage Spikes}

\begin{figure*}[t!]
  \centering
  \includegraphics[width=0.76\textwidth]{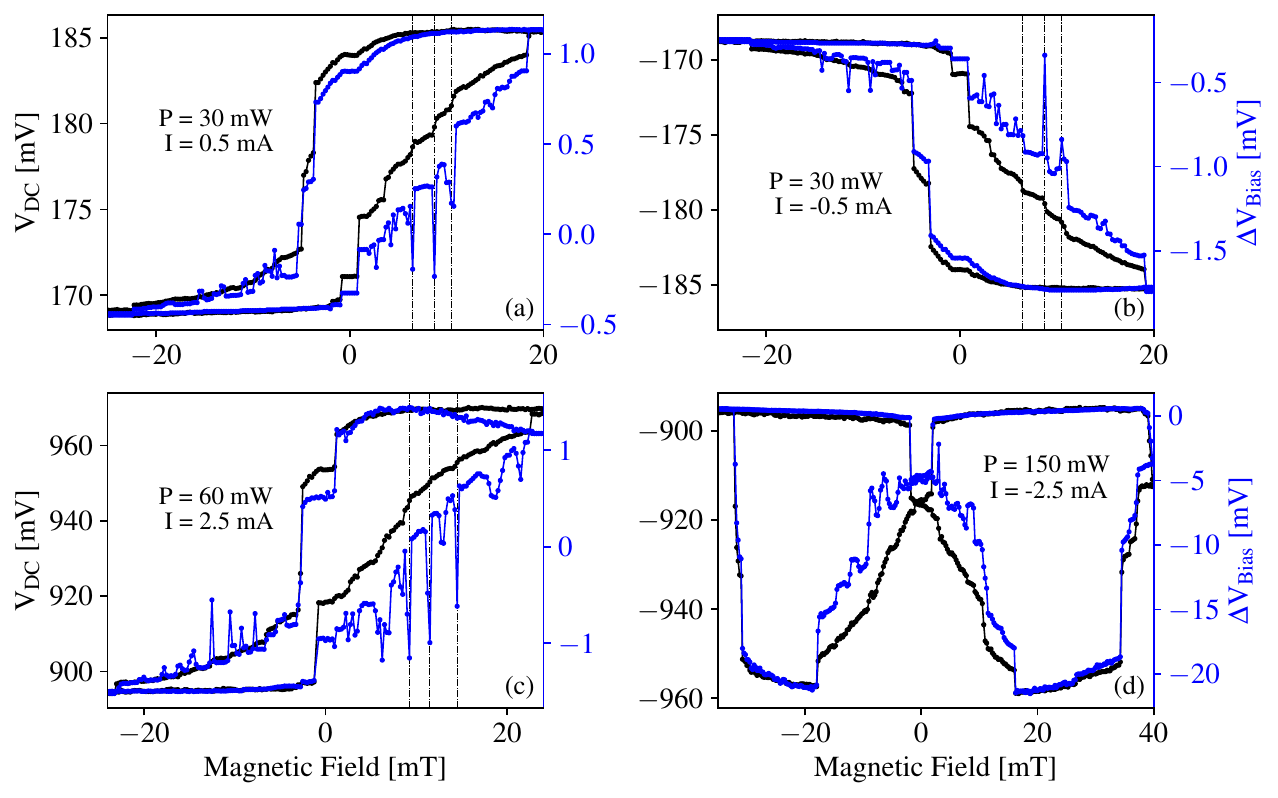}
  \caption{The dependence of the $\mathrm{\Delta V_{Bias}}$ and $\mathrm{\Delta V_{DC}}$ on the external magnetic field for different laser powers and bias currents in a nanopillar with a d = 800nm.}
\label{Fig2SM}
\end{figure*}

\begin{figure*}[ht!]
  \centering
  \includegraphics[width=0.76\textwidth]{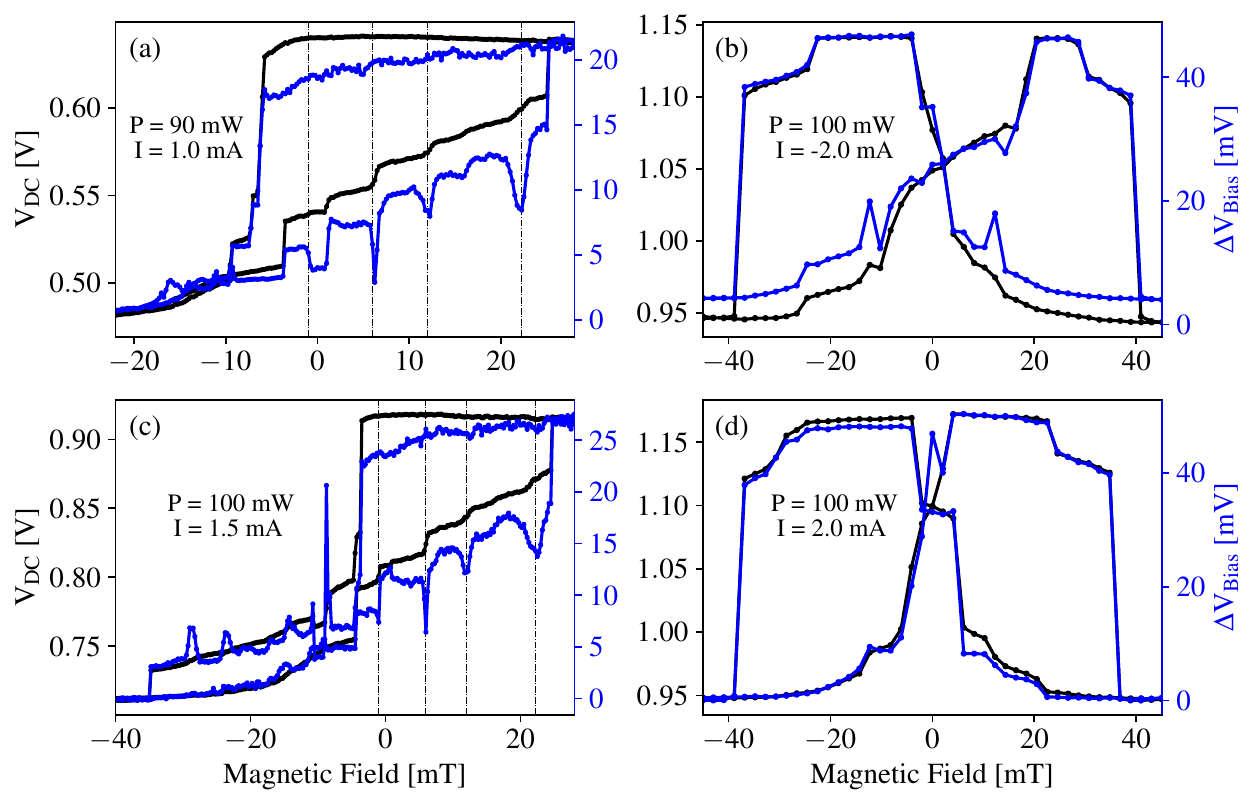}
  \caption{The dependence of the $\mathrm{\Delta V_{Bias}}$ and $\mathrm{\Delta V_{DC}}$ on the external magnetic field for different laser powers and bias currents in a nanopillar with a d = 300 nm.}
\label{Fig3SM}
\end{figure*}

Fig. \ref{Fig2SM} illustrates the changes in the DC and AC components of the output voltage during the magnetization sweep for an STNO with an 800 nm nanopillar diameter. For all devices, the switching of $\mathrm{V_\text{AC}}$ and $\mathrm{V_\text{DC}}$ occurs at identical magnetic field values, although their voltage magnitudes differ, highlighting the potential for multi-state operation in the system. Notably, under the configuration with 150 mW laser power and a bias current of -2.5~mA, we observe not only a significantly larger thermovoltage but also an intriguing double-switching phenomenon in both the DC and AC voltage components. This novel behavior reveals the intricate interplay between thermal gradients, bias currents, and magnetization dynamics, shedding light on the potential for advanced functionalities in spintronic devices. Furthermore, the TMR, i.e., the relative change of resistance upon magnetization reversal, is defined as $\mathrm{TMR=\frac{R_{AP}-R_P}{min(|R_P|, |R_{ap}|)}}$ \cite{jha2023interface, sadat2023enhancing}. The calculated TMR values for the devices under different laser power and bias current conditions, as presented in Fig.~\ref{Fig2SM} (a)–(d), are approximately 9.7$\%$, 9.7$\%$, 8.3$\%$, and 6.4$\%$, respectively. A similar double-switching behavior is observed in Fig. \ref{Fig3SM} for the STNO device with a nanopillar diameter of 300 nm, further supporting the universality of this effect in STNOs. In particular, when P = 100 mW and I = 1.5 mA, the AC and DC components of the output voltages exhibit asymmetric hysteresis behavior, which is equivalent to generating voltage states. By increasing the current, as shown in (b), we observe the emergence of double-switching behavior. Furthermore, the TMR ratio, based on the total resistance of the device without subtracting effects of the pad, is 33$\%$ for Fig.~\ref{Fig3SM} (a), 21$\%$ for (b), 28$\%$ for (c), and 25$\%$ for (d).

In Figs. \ref{Fig2SM} and \ref{Fig3SM}, vertical lines indicate the magnetic field values corresponding to Barkhausen jumps observed in the DC voltage or resistance of the devices. Notably, these points align with pronounced jumps in the thermovoltage, revealing a strong correlation between domain dynamics, vortex state transitions, and thermoelectric responses in the material. This suggests that thermovoltage measurements can serve as a more sensitive probe of domain motion and vortex dynamics, uncovering aspects of magnetization behavior that may be missed by traditional electrical measurements. These insights not only advance our understanding of fundamental spintronics phenomena, but also position thermovoltage-based measurements as a promising tool for applications in reservoir computing, where complex nonlinear dynamics and memory effects are critical for efficient information processing.

\begin{figure*}[ht!]
  \centering
  \includegraphics[width=0.99\textwidth]{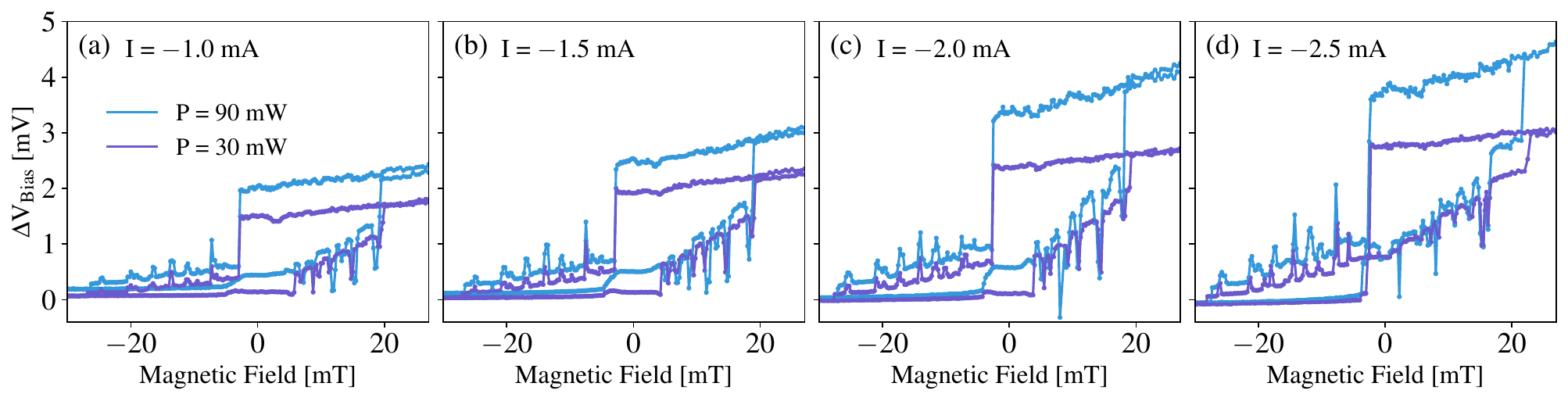}
  \caption{The dependence of the $\mathrm{\Delta V_{Bias}}$ on the external magnetic field for different laser powers and bias currents in a nanopillar with a d = 600 nm.}
\label{Fig4SM}
\end{figure*}

To further illustrate the presence of thermovoltage spikes, a detailed figure for a device with d = 600 nm is presented in Fig. \ref{Fig4SM}. The thermovoltage spikes align with step-like features in the DC voltage, indicating a strong correlation with abrupt magnetization transitions and vortex core dynamics. This highlights the thermovoltage’s exceptional sensitivity to magnetization changes, offering insights that go beyond conventional DC measurements. These spikes present promising opportunities for applications in neuromorphic computing, high-resolution magnetic sensing, and reservoir computing, where nonlinear dynamics and spiking behavior are essential for advanced information processing.

\bibliography{2_SM.bib}